\documentclass[11pt,twoside]{article}

\usepackage{asp2006}
\usepackage{epsf}
\usepackage{graphicx}

\begin{document}
\title{Accretion Disk Winds in AM CVn Binaries -- a Monte Carlo Approach}   %%% Fill in title
\author{D.-J\@. Kusterer, T\@. Nagel, and K\@. Werner}   %%% Fill in author names
%\aindex{Kusterer, D.-J.|bb}
%\aindex{Nagel, T.}
%\aindex{Werner, K.}

\affil{Institute for Astronomy and Astrophysics, 
           Kepler Center for Astro and Particle Physics, 
           Eberhard Karls University, T\"ubingen, Germany}    %%% Fill in author affiliations

\begin{abstract} %%% Abstract to run on from here.
AM CVn systems are interacting binaries similar to cataclysmic variables (CVs), but more compact with orbital periods of less than 80 minutes. The primary is a white dwarf, whereas the nature of the secondary is not completely clear, yet. Abundances and composition of the outer layer of the secondary can be found by analysis of the accretion disk (presented by Nagel et al. these proceedings). Spectra from high-state AM CVn systems do not only show typical signatures of accretion disks, but also P Cygni line profiles, a sign of outflow being present in the system. Here we present the first quantitative spectral analysis of an accretion-disk wind in AM CVn systems. Emergent wind spectra are modeled with our 3-D Monte Carlo radiative transfer code WOMPAT. We show that P Cygni profiles can be reproduced with our wind models.  
\end{abstract}

%%% MAIN BODY OF TEXT GOES HERE. CONSULT "INSTRUCTIONS FOR AUTHORS USING
%%% LATEX2E MARKUP", SECTIONS 2.3-2.6 FOR HELP WITH EQUATIONS, FIGURES,
%%% AND TABLES.

%\section{}   %%% Top level section head (remove "%" symbol)
%\subsection{}   %%% Second level section head (remove "%" symbol)
%\subsubsection{}   %%% Lowest level section head (remove "%" symbol)
%\section*{}    %%% Unnumbered top level section head (remove "%" symbol)
%\subsection*{}   %%% Unnumbered second level section head (remove "%" symbol)
The first detection of outflows in CVs dates back to the late 1970ies and early 1980ies when blueshifted absorption troughs and P~Cygni profiles in ultraviolet (UV) resonance lines were discovered  \citep{DJK1978Natur.275..385H,DJK1982ApJ...260..716C}. More recent work shows that outflowing material has a strong influence on observations, not only in the ``classical'' UV wind lines, but also in other features. Not only in CVs, but also in their ``smaller twins'' the AM CVn systems, signatures of biconical outflow are found. As in CVs such an accretion-disk wind is mainly found in high-state systems, a complete model of such systems therefore has to include an outflow.

\begin{figure}[t!]%hbp]
\begin{center}
\includegraphics[angle=270,width=0.95\textwidth]{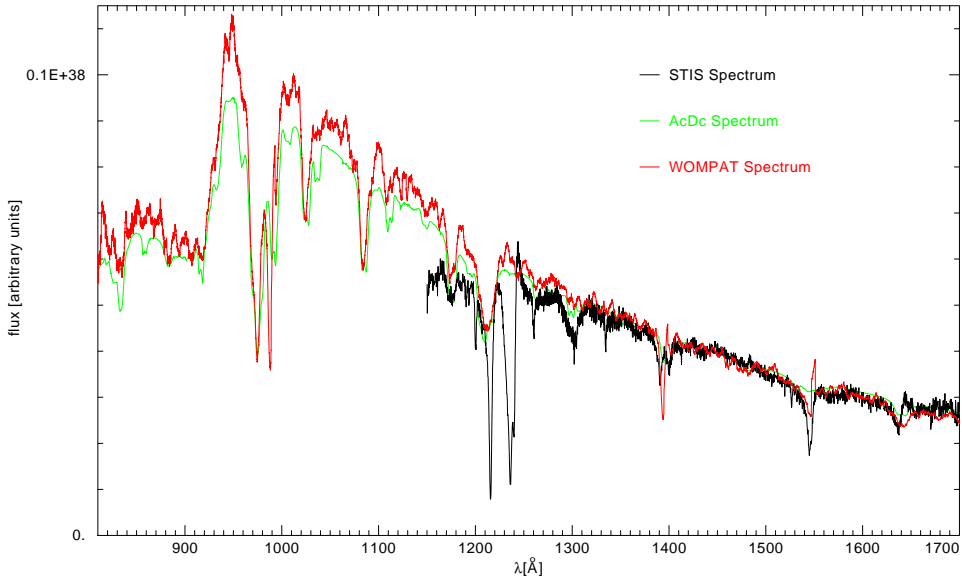}
\caption{STIS spectrum of AM CVn overlaid with a theoretical disc spectrum calculated with AcDc \citep{DJK2004A&A...428..109N} and a Monte Carlo accretion disk wind spectrum calculated with our new code WOMPAT.}
\label{DJKspec}
\end{center}
\end{figure}

Within the T\"ubingen group a code for modeling NLTE accretion-disk atmospheres was developed and successfully used to model spectra of CVs, AM CVn systems and ultracompact X-ray binaries \citep{DJK2004A&A...428..109N,DJK2006A&A...450..725W}. Our goal is to develop a whole package with which we are able to model a complete CV including the accretion disk, the white dwarf and the outflow. Here our first results with disk, a blackbody WD and wind for the prototype AM CVn are presented. We implemented a kinematical biconical wind model by \cite{DJK1993ApJ...409..372S}. Monte Carlo (MC) techniques are used to do the radiative transfer in this 3-D wind structure. Photon packets, which represent a monochromatic family of photons, are created and then followed through the wind. All parts of a photon packet's life, like the creation, interactions, new directions of flight are determined via probabilities. Optical depths are acquired by numerical integration of local opacities along the photon's line of flight. No Sobolev approximation is needed. Furthermore line opacities can be calculated with either Doppler or Stark broadening. Typical P~Cygni wind lines, such as the C{\sc iv} 1550~\AA\ resonance line (Fig.~\ref{DJKspec}) are reproduced by our model. Despite the obvious difference in chemical composition and size of the accretion disks in AM CVn and normal, hydrogen rich, CVs, the wind lines in both cases are quite similar. This means that the dependence of the wind mechanism on these parameters has to be relatively small.
Figure~\ref{DJKspec} shows a STIS spectrum of AM CVn itself, which shows several weak P Cygni features. Also included in Fig.~\ref{DJKspec} is a spectrum calculated with WOMPAT for an AM CVn  parameter set with $M_{\mathrm{WD}} = 0.6\, \mathrm{M}_{\odot}$ and an inclination angle of $40^{\circ}$. No parameter space exploration has been done for this, therefore it is just to be taken as a very rough guidance for future progress. As a source of radiation for the photon packets a blackbody white dwarf with $T_{ \mathrm{eff}} = 18,000\, \mathrm{K}$ and the accretion disc were used. The accretion disk input consisted of twelve rings calculated with {\sc AcDc} \citep{DJK2004A&A...428..109N} and further blackbody rings to fill up the gap to the outer edge of the disk. The combined theoretical spectrum of these twelve {\sc AcDc} rings is shown in Fig.~\ref{DJKspec}, as well. It is clearly seen that the disk alone cannot produce a C{\sc iv} 1550 \AA\ P Cygni shape.\\ 

\acknowledgements %%% Text of acknowledgements runs on after this command.
This work is supported by DFG grant We 1312/37-1.

\end{document}